\documentclass{ws-procs9x6}

\newcommand{\vk}{\mbox {\boldmath $k$\unboldmath}}
\newcommand{\vq}{\mbox {\boldmath $q$\unboldmath}}

\begin{document}

\title{Beyond the known baryon resonances}

\author{B. SAGHAI}

\address{DAPNIA, CEA/Saclay, 91191 Gif-sur-Yvette, France\\
E-mail: bsaghai@cea.fr}

\author{Z. LI}

\address{Physics Department, Peking University,
Beijing 100871, P.R. China}


\maketitle
\abstracts{
$\eta$ photoproduction data on the proton up to 
$E_\gamma ^{lab} \approx$2 GeV are interpreted within a chiral constituent 
quark formalism, which embodies all known three and four star resonances. 
This study confirms the need for a new $S_{11}$ resonance, with
M=1.780 GeV and $\Gamma$=280 MeV, already
introduced in investigating lower energy data.}
\section{Introduction}
Recent data\cite{CLAS02,Graal02,Jlab-e,pol,Mainz} on the electromagnetic 
production of the $\eta$ meson off the proton constitute an exciting challenge 
for phenomenologists\cite{LS-3,LS-2,LS-1,QSL02,WT02a,WT02b}.
This isospin pure process, dominated at low energies by a single nucleon
resonance, offers appealing features not only for pinning down the 
reaction mechanism and the extraction of the fundamental $\eta N$
coupling constant, but also for search of new nucleon resonances\cite{LS-3,LS-2}.
Then, such experimental and theoretical efforts are expected to
test various QCD-inspired approaches\cite{Review,GSV} in the hadron spectroscopy 
realm, allowing strong test of the underlying concepts.

In this note, we investigate the reaction $\gamma p \to \eta p$, and 
report on the results of a chiral constituent 
quark formalism\cite{LS-1,zpl97}, which embodies\cite{LS-2} 
the configuration mixing phenomenon\cite{IK,kar02}, and the related
$SU(6)\otimes O(3)$ symmetry breaking effects\cite{LS-2,LS-1}.
%
%
\section{Theoretical Frame}
The chiral constituent quark approach for meson photoproduction 
is based on the low energy QCD Lagrangian\cite{MANOHAR},
with four components for the photoproduction of
pseudoscalar mesons.
The first one is a seagull term, generated by the gauge 
transformation of the axial vector $A_{\mu}$ in the QCD Lagrangian.
The second and the third terms correspond to the {\it s-} and {\it u-}channels,
respectively. 
The forth term is the {\it t-}channel contribution and contains two parts: 
{\it i)} charged meson exchanges which are proportional to the charge of outgoing 
mesons and thus do not contribute to the process $\gamma N\to \eta N$;  
{\it ii)} $\rho$ and $\omega$ exchange in the $\eta$ production which are 
excluded here due to the duality hypothesis\cite{LS-2}.

The contributions from  the {\it s-}channel resonances can be written as
\begin{eqnarray}\label{eq:MR}
{\mathcal M}_{N^*}=\frac {2M_{N^*}}{s-M_{N^*} \big [ M_{N^*}
-i\Gamma(q) \big ]}
e^{-\frac {{k}^2+{q}^2}{6\alpha^2_{ho}}}{\mathcal A}_{N^*},
\end{eqnarray}
with  $k=|\vk|$ and $q=|\vq|$ the momenta of the incoming photon 
and the outgoing meson respectively, $\sqrt {s}$ the total energy of 
the system, $e^{- {({k}^2+{q}^2)}/{6\alpha^2_{ho}}}$ is a form factor 
in the harmonic oscillator basis with the parameter $\alpha^2_{ho}$ 
related to the harmonic oscillator strength in the wave-function, 
and $M_{N^*}$ and $\Gamma(q)$ are the mass and the total width of 
the resonance, respectively.  The amplitudes ${\mathcal A}_{N^*}$ 
are divided into two parts\cite{zpl97}: the contribution 
from each resonance below 2 GeV, the transition amplitudes of which 
have been translated into the standard CGLN amplitudes in the harmonic 
oscillator basis, and the contributions from the resonances above 2 GeV
treated as degenerate, since little experimental information is available
on those resonances.

The contributions from each resonance to $\eta$
photoproduction is determined by introducing\cite{LS-1} a new set of 
parameters $C_{{N^*}}$ and the 
${\mathcal A}_{N^*} \to C_{N^*} {\mathcal A}_{N^*}$ 
substitution rule for the 
amplitudes, so that 
${\mathcal M}_{N^*}^{exp} = C^2_{N^*} {\mathcal M}_{N^*}^{qm}$,
where ${\mathcal M}_{N^*}^{exp}$ is the experimental value of 
the observable, and ${\mathcal M}_{N^*}^{qm}$ is calculated in the 
quark model\cite{zpl97}. 
The $SU(6)\otimes O(3)$ symmetry predicts
$C_{N^*}$~=~0 for ${S_{11}(1650)} $, ${D_{13}(1700)}$, and 
${D_{15}(1675)} $ resonances, and $C_{N^*}$~=~1 for other relevant
resonances: 
$S_{11}(1535)$, $P_{11}(1440)$,$P_{11}(1710)$, $P_{13}(1720)$,
$P_{13}(1900)$, $D_{13}(1520)$, $F_{15}(1680)$, and $F_{15}(2000)$.   
Thus, the coefficients $C_{{N^*}}$ give a measure of the discrepancies 
between 
the theoretical results and the data and show the extent 
to which the $SU(6)\otimes O(3)$ symmetry is broken in the process 
investigated here.

One of the main reasons that the $SU(6)\otimes O(3)$ symmetry is
broken is due to the configuration mixings caused by the one gluon
exchange\cite{IK,kar02}. 
Here, the most relevant configuration mixings are those of the
two $S_{11}$ and the two $D_{13}$ states around 1.5 to 1.7 GeV. The 
configuration mixings can be expressed in terms of the mixing angle
between the two $SU(6)\otimes O(3)$ states with the total quark spin 
1/2 and 3/2.  
\section{Results and Discussion}
Using our approach and the MINUIT minimization code from the 
CERN Library, we have fitted all $\approx$ 650 data points
from recent measurements for both differential 
cross-sections\cite{CLAS02,Graal02,Mainz}
and single polarization asymmetries\cite{pol}.
The adjustable parameters of our models are one $SU(6)\otimes O(3)$ symmetry
breaking strength coefficient ($C_{N^*}$) per resonance, except for the resonances
$S_{11}(1535)$ and $S_{11}(1650)$ on the one hand, and 
$D_{13}(1520)$ $D_{13} (1700)$ on the other hand, for which we introduce
the configuration mixing angles $\theta _{S}$ and $\theta _{D}$.

The first model includes explicitly all 
eleven known relevant resonances, mentioned above, with mass below 2 GeV, 
and the contributions from the known excited resonances above 2 GeV for a 
given parity. assumed to be degenerate and hence written in a 
compact form\cite{zpl97}.

In Fig.~1, we compare this model (dashed curves) to the data at nine
incident photon energies. As shown in our earlier works\cite{LS-2,LS-1},
such a model reproduces correctly the data at low energies
($E_\gamma ^{lab} \le$ 1 GeV). Above, the model misses the data. 
A possible reason for these theory/data discrepancies could be that
some yet unknown resonances contribute to the reaction mechanism.
We have investigated possible r\^ole played by extra $S_{11}$, $P_{11}$, and $P_{13}$
resonances, with three free parameters 
(namely the resonance mass, width, and strength) in each case.

By far, the most significant improvement was obtained by a third
$S_{11}$ resonance, with the extracted values M=1.780 GeV 
and $\Gamma$=280 MeV. The configuration mixing angles came out to be 
$\theta _{S}$=12$^\circ$ and $\theta _{D}$=-35$^\circ$, in agreement
with the Isgur-Karl model~\cite{IK} and by large-$N_c$ approaches~\cite{Nc}.
  
The outcome of this latter model is depicted in Fig.~1 (full curve)
and shows very reasonable agreement with the data, improving the reduced
$\chi^2$, on the complete data-base, by more than a factor of 2.
%
\begin{figure}[b!]
\epsfysize=8.9cm 
\centerline{\epsfxsize=4.7in\epsfbox{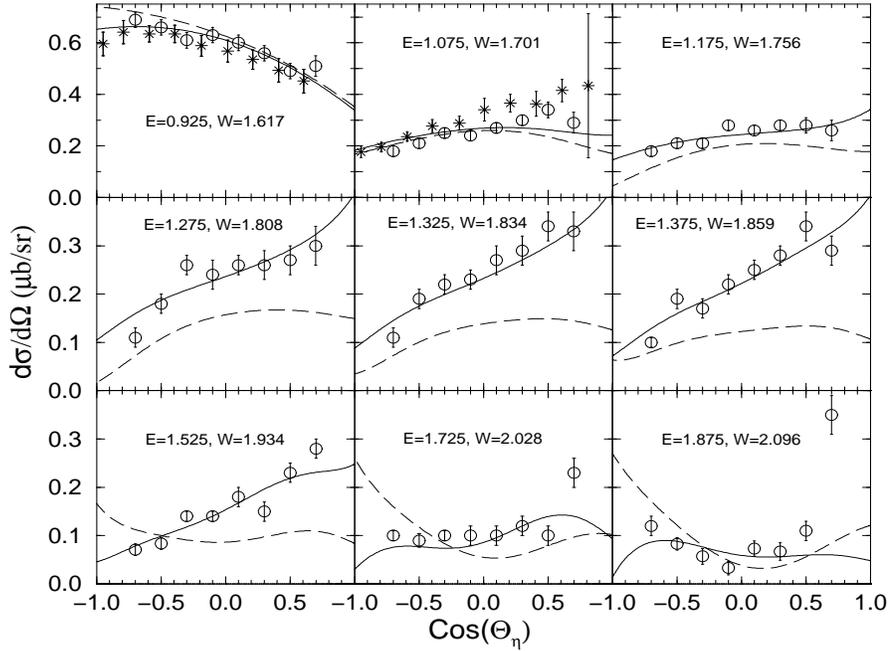}}   
\caption{Differential cross section for the process 
$\gamma p \to \eta p$: angular distribution at nine
incident photon energies ($E_{\gamma}^{lab}$), with the
corresponding total center-of-mass energy (W) also given;
units are in GeV.
The dashed curves are from the model embodying all known
three and four star resonances. The full curves show the
model including, in addition, a new $S_{11}$ resonance,
with M=1.780 GeV and  $\Gamma$=280 MeV.
CLAS (circles) and GRAAL (stars) data are
from Refs. [1] and [2], respectively.\label{inter}
}
\end{figure}
%
The extracted values for the mass and width of a new $S_{11}$ are close 
to those predicted by the authors of Ref.\cite{LW96}
(M=1.712 GeV and $\Gamma$=184 MeV), 
and our previous findings\cite{LS-2}.
Moreover, for the one star $S_{11}(2090)$ resonance\cite{PDG}, the 
Zagreb group
coupled channel analysis\cite{Zagreb} produces the following values
M = 1.792 $\pm$ 0.023 GeV and $\Gamma$ = 360 $\pm$ 49 MeV.
The BES Collaboration
reported\cite{BES} on the measurements of the
$J/\psi \to p \overline{p} \eta$ decay channel. 
In the latter work, a partial wave analysis
leads to the extraction of the mass and width of the 
$S_{11}(1535)$ and $S_{11}(1650)$ resonances, and the authors find  
indications for an extra resonance with 
M = 1.800 $\pm$ 0.040 GeV, and $\Gamma$ = 165$^{+165}_{-85}$ MeV.
A very recent work\cite{GSV} based on the hypercentral constituent 
quark model, and presented during this workshop,
predicts a missing $S_{11}$ resonance with M=1.861 GeV.
Finally, a self-consistent analysis of pion scattering and photoproduction
within a coupled channel formalism, indicates\cite{che02} the 
existence of a third $S_{11}$ resonance with M =1.803 $\pm$ 0.007 GeV.

The main shortcoming of our model concerns the deviations between
theory and very forward data at highest energies (Fig.~1), indicating
the need for a more careful treatment of higher mass resonances or,
equivalently, the introduction of the t-channel contributions\cite{WT02b}.

Finally, given the quality of the data, coupled-channel effects
studied\cite{CC} in the associated strangeness photoproduction
sector, have to be extended to the $\eta$ photoproduction process. 

In summary, results presented in this note and those reported
in the quoted works, allow us to confidently conclude that the existence
of a third $S_{11}$ resonance with M$\approx$1.8 GeV is being
established. Our preliminary results on the double polarization
asymmetries show that some of those observables can provide us
with strong criteria on the issues of new resonances.
Such measurements will, hopefully, be performed in the near future 
in GRAAL, JLAB, and/or Spring\_8.

\bigskip

One of us (BS) wishes to thank the organizers for their kind invitation 
to this very stimulating workshop.

\end{document}